\documentclass[aps,pre,superscriptaddress,preprint,amsmath,amsfonts,amssymb,a4paper]{revtex4-1}
\usepackage{graphicx}
\usepackage{hyperref}
\newcommand{\beq}{\begin{eqnarray}}
\newcommand{\eeq}{\end{eqnarray}}

\begin{document}
\title{New representations of $\pi$ and Dirac delta using the nonextensive-statistical-mechanics $q$-exponential function}

\author{M. Jauregui}
\affiliation{Centro Brasileiro de Pesquisas Fisicas and National Institute of Science and Technology for Complex Systems, Rua Xavier Sigaud 150, 22290-180 Rio de Janeiro, Brazil}
\author{C. Tsallis}
\affiliation{Centro Brasileiro de Pesquisas Fisicas and National Institute of Science and Technology for Complex Systems, Rua Xavier Sigaud 150, 22290-180 Rio de Janeiro, Brazil}
\affiliation{Santa Fe Institute, 1399 Hyde Park Road, Santa Fe, NM 87501, USA}

\begin{abstract}
We present a generalization of the representation in plane waves of Dirac delta, $\delta(x)=\frac{1}{2\pi}\int_{-\infty}^\infty e^{-ikx}\,dk$, namely $\delta(x)=\frac{2-q}{2\pi}\int_{-\infty}^\infty e_q^{-ikx}\,dk$, using the nonextensive-statistical-mechanics $q$-exponential function, $e_q^{ix}\equiv[1+(1-q)ix]^{1/(1-q)}$ with $e_1^{ix}\equiv e^{ix}$, being $x$ any real number, for real values of $q$ within the interval $[1,2[$. Concomitantly, with the development of these new representations of Dirac delta, we also present two new families of representations of the transcendental number $\pi$. Incidentally, we remark that the $q$-plane wave form which emerges, namely, $e_q^{ikx}$, is normalizable for $1<q<3$, in contrast with the standard one, $e^{ikx}$, which is not.
\end{abstract}

\maketitle

\section{Introduction}
Dirac delta is a distribution that is used in almost all branches of physics. Various representations of it have been discovered along the time. For example, it can be represented as a limit of a Gaussian or as a linear combination of plane waves, being the last one strongly related to the Fourier transform (FT), as we will show later.

Dirac delta, $\delta(x)$, obeys the following fundamental property:
\beq
\int_{-\infty}^\infty f(x)\delta(x)\,dx=f(0)\,,\label{delta}
\eeq
where $f:\mathbb{R}\to\mathbb{C}$ is a well-behaved function. From the equation above, we can see that if $f(x)=1\;\forall x\in\mathbb{R}$, we get the normalization condition
\beq
\int_{-\infty}^\infty \delta(x)\,dx&=&1\,.\label{deltanormal}
\eeq
Also, choosing $f(x)=f(0)e^{ikx}$ in \eqref{delta}, we obtain  
\beq
\int_{-\infty}^\infty \delta(x) e^{ikx}\,dx=1\,,\label{delta2}
\eeq
i.e., the FT of $\delta(x)$ equals one. Therefore, using the expression of the inverse FT we obtain the following representation of Dirac delta:
\beq\label{deltafourier}
\delta(x)=\frac{1}{2\pi}\int_{-\infty}^\infty e^{-ikx}\,dk\,,
\eeq
which can be interpreted as a linear combination of plane waves. We can rewrite the expression above as
\beq\label{intlim}
\delta(x)=\frac{1}{2\pi}\lim_{L\to\infty}\int_{-L}^L e^{-ikx}\,dk\,,
\eeq
then, Dirac delta also can be represented as the following improper limit:
\beq\label{deltalim}
\delta(x)=\lim_{L\to\infty}\frac{\sin(Lx)}{\pi x}\,.
\eeq

In 1988, a possible generalization of Boltzmann-Gibbs statistical mechanics was proposed\cite{Tsallis1988}. This new theory, sometimes referred to as {\it nonextensive statistical mechanics} \cite{books}, has been satisfactorily applied to handle a large number of physical phenomena (usually, metastable or quasi-stationary states of  systems that are not consistent with the ergodic hypothesis; for example, systems in which long-range interactions or strong-correlations exist)\cite{UpadhyayaRieuGlazierSawada2001,DanielsBeckBodenschatz2004,ArevaloGarcimartinMaza2007a,DouglasBergaminiRenzoni2006,LiuGoree2008,DeVoe2009,Borland2002a,Queiros2005,BurlagaVinas2005,BurlagaNess2009,BakarTirnakli2009,CarusoPluchinoLatoraVinciguerraRapisarda2007,CarvalhoSilvaNascimentoMedeiros2008,PickupCywinskiPappasFaragoFouquet2009,CMS2010}.
Furthermore, the elaboration of nonextensive statistical mechanics required the generalization of some mathematical functions (exponential, logarithm, etc.), operators (sum, product, Fourier transform, etc.) and theorems (central limit theorem)\cite{UmarovTsallisSteinberg2008}. Particularly, the generalization of the exponential function, namely, the $q$-exponential function is defined by
\beq
e_q^{x}\equiv[1+(1-q)x]_+^\frac{1}{1-q}\qquad\left(e_1^{x}\equiv e^{x}\right)\,,
\eeq
for any $x\in\mathbb{R}$, where the symbol $[y]_+$ means that $[y]_+=y$, if $y\geq0$, and $[y]_+=0$ if $y<0$. For pure imaginary $ix$, $e_q^{ix}$ can be defined to be the principal value of
\beq \label{def}
e_q^{ix}\equiv[1+(1-q)ix]^\frac{1}{1-q}\qquad\left(e_1^{ix}\equiv e^{ix}\right)\,.
\eeq

The main purpose of the present paper is to generalize the representation in plane waves of Dirac delta, introduced in equation \eqref{deltafourier}, using the $q$-exponential function defined above.
\section{Representation of Dirac delta in $q$-plane waves }

\subsection{Proposition}
Let us introduce the following quantity:
\beq\label{qdelta}
\delta_q(x)\equiv\frac{1}{c(q)}\int_{-\infty}^\infty e_q^{-i\xi x}\,d\xi\,\text{ with }q\in[1,2[ \,,
\eeq
which can be interpreted as a linear combination of $q$-plane waves, where $c(q)$ is a constant that may depend on $q$. We intend to show later that $\delta_q(x)=\delta(x)$ for all $1 \le q<2$.

Analogously to \eqref{intlim}, we may write
\beq\label{qdelta2}
\delta_q(x)=\frac{1}{c(q)} \lim_{\Lambda \to\infty}\int_{-\Lambda}^\Lambda e_q^{-i\xi x}\,d\xi\,\text{ with }q\in[1,2[ \,,
\eeq
therefore, by integrating, we can represent $\delta_q(x)$ as the following improper limit:
\beq\label{qdeltalim}
\delta_q(x)=\frac{2}{(2-q)c(q)}\lim_{\Lambda\to\infty}\frac{\sin\left\{\frac{2-q}{q-1}\arctan[(q-1)\Lambda x]\right\}}{x\left[1+(q-1)^2\Lambda^2x^2\right]^\frac{2-q}{2(q-1)}}\,\text{ with }q\in]1,2[  \,.
\eeq

\subsection{The normalization constant $1/c(q)$ and the transcendental number $\pi$}
The constant $c(q)$ must be equal to $2\pi$ at the limit $q\to1^+$. Furthermore, $c(q)$ can be found from the normalization condition \eqref{deltanormal}. Thus, we have
\beq
c(q)=\frac{2}{(2-q)}\lim_{\Lambda\to\infty}\int_{-\infty}^\infty\frac{\sin\left\{\frac{2-q}{q-1}\arctan[(q-1)\Lambda x]\right\}}{x\left[1+(q-1)^2\Lambda^2x^2\right]^\frac{2-q}{2(q-1)}}\,dx\,.
\eeq
Using the change of variables $z=(q-1)\Lambda x$ we obtain
\beq
c(q)=\frac{2}{(2-q)}\lim_{\Lambda\to\infty}\int_{-\infty}^\infty\frac{\sin\left(\frac{2-q}{q-1}\arctan z\right)}{z(1+z^2)^\frac{2-q}{2(q-1)}}\,dz\,.
\eeq
As the integral does not depend on $\Lambda$, the limit symbol can be omitted. Therefore, we can write
\beq\label{cqint}
c(q)=\frac{2}{(2-q)}\int_{-\infty}^\infty\frac{\sin[2\alpha(q)\arctan z]}{z\left(1+z^2\right)^{\alpha(q)}}\,dz\,,
\eeq
where
\beq\label{alphaq}
\alpha(q) \equiv \frac{2-q}{2(q-1)}\,.
\eeq
We easily verify that $\alpha:]1,2[\subset\mathbb{R}\to\mathbb{R^+}$ is a monotonically decreasing function of $q$.

In order to solve analytically the integral in \eqref{cqint}, let us restrict to integer or half-integer values for $\alpha(q)$, more precisely, $1/2,1,3/2,\ldots$. This implies that $q$ will be allowed to assume just certain rational values within the interval $]1,2[$, namely $q=3/2,4/3,5/4,\ldots$. Using the change of variables $z=\tan\theta$ in Eq. \eqref{cqint}, we obtain
\beq
c(q)=\frac{4}{2-q}\int_0^{\pi/2}\frac{\sin[2\alpha(q)\theta](\cos\theta)^{2\alpha(q)-1}}{\sin\theta}\,d\theta \,.
\eeq
By using now the relation \eqref{trigidentity} proved in the Appendix, the expression above yields
\beq\label{cqbeta}
c(q)=\frac{4}{2-q}\sum_{k=0}^{\left\lfloor\alpha(q)+\frac{1}{2}\right\rfloor-1}(-1)^k\binom{2\alpha(q)}{2k+1}\int_0^{\pi/2} d\theta\,(\cos\theta)^{4\alpha(q)-2k-2}(\sin\theta)^{2k}\,.
\eeq
We remind that the Beta function, $B(x,y)$, is defined by
\beq
B(x,y)\equiv\int_o^{\pi/2}d\phi\,2(\cos\phi)^{2x-1}(\sin\phi)^{2y-1}\,\text{ with }x>0\text{ and }y>0\,,
\eeq
which is related with Gamma function by
\beq
B(x,y)=\frac{\Gamma(x)\Gamma(y)}{\Gamma(x+y)}\,.
\eeq
Therefore, using the expressions of Beta function shown above, Eq. \eqref{cqbeta} can be written as
\beq\label{cqgamma}
c(q)=\frac{4\alpha(q)}{2-q}\sum_{k=0}^{\left\lfloor\alpha(q)+\frac{1}{2}\right\rfloor-1}(-1)^k\frac{\Gamma\left(2\alpha(q)-k-\frac{1}{2}\right)\Gamma\left(k+\frac{1}{2}\right)}{\Gamma(2k+2)\Gamma(2\alpha(q)-2k)}\,.
\eeq
Let us rewrite now the expression above as
\beq\label{cqpi}
c(q)=\frac{2}{2-q}S_{n_q}\,,
\eeq
where
\beq
S_{n_q}\equiv n_q\sum_{k=0}^{\left\lfloor\frac{n_q+1}{2}\right\rfloor-1}(-1)^k\frac{\Gamma\left(n_q-k-\frac{1}{2}\right)\Gamma\left(k+\frac{1}{2}\right)}{\Gamma(2k+2)\Gamma(n_q-2k)}\,\text{ with }n_q \equiv 2\alpha(q)\in\mathbb{N}\,.
\eeq
When $n_q=1$ (which corresponds to $q=3/2$) we obtain straightforwardly that $S_1=\pi$. Also, it is straightforward to verify that $S_2,S_3,S_4$ are equal to $\pi$. Using a symbolic computation software, we also verified that from $n_q=1$ to $n_q=5000$ ($q=5002/5001$), $S_{n_q}=\pi$. Hence we state the following hypothesis:
\beq\label{pigamma}
\pi=n\sum_{k=0}^{\left\lfloor\frac{n+1}{2}\right\rfloor-1}(-1)^k\frac{\Gamma\left(n-k-\frac{1}{2}\right)\Gamma\left(k+\frac{1}{2}\right)}{\Gamma(2k+2)\Gamma(n-2k)} \,\quad\forall n\in\mathbb{N}\,.
\eeq
We thus found a countable infinite family of representations of the transcendental number $\pi$ (see also\cite{hypergeometricseriesbook}).

\begin{figure}[htp]
\centering
\includegraphics[width=0.5\textwidth,keepaspectratio]{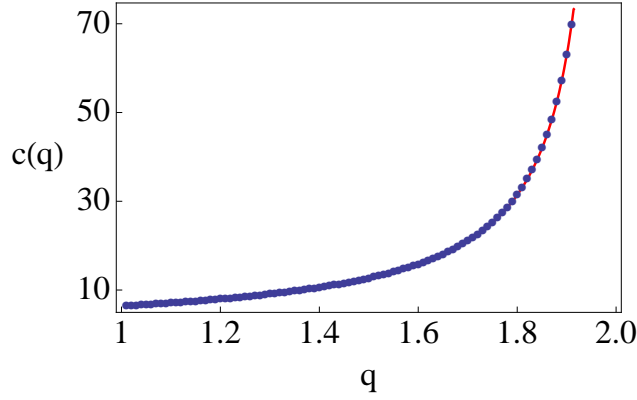}
\caption{The blue dots were numerically obtained using expression \eqref{cqint}, whereas the red continuous curve is the plot of $c(q)$ given by equation \eqref{cq}.}
\label{fcq}
\end{figure}

Using relation \eqref{pigamma} in \eqref{cqpi}, the expression of $c(q)$ becomes
\beq\label{cq}
c(q)=\frac{2\pi}{2-q}\,.
\eeq
In addition to the above, this relation has been checked numerically to be correct not only for certain rational values of $q$ within the interval $[1,2[$, but for all real numbers within that interval (see Fig. \ref{fcq}). Therefore, we conjecture that the integral which appears in \eqref{cqint} equals $\pi$ for {\it any} value of $q$ within that interval. Consistently, we obtain another infinite family of representations of the number $\pi$, namely
\beq\label{piint}
\pi=\int_{-\infty}^\infty\frac{\sin(2r\arctan z)}{z\left(1+z^2\right)^r}\,dz\,\quad\forall r\in\mathbb{R}^+\,.
\eeq
This family is non countable and contains Eq. (\ref{pigamma}) as a particular case.

Finally, expressions \eqref{qdelta} and \eqref{qdeltalim} of $\delta_q(x)$ become respectively
\beq\label{qdeltafinal}
\delta_q(x)=\frac{2-q}{2\pi}\int_{-\infty}^\infty e_q^{-i\xi x}\,d\xi\,\text{ with }q\in[1,2[
\eeq
and 
\beq
\delta_q(x)=\lim_{\Lambda\to\infty}\frac{\sin\left\{\frac{2-q}{q-1}\arctan[(q-1)\Lambda x]\right\}}{\pi x\left[1+(q-1)^2\Lambda^2x^2\right]^\frac{2-q}{2(q-1)}}\,\text{ with }q\in]1,2[\,.\label{qdeltalimfinal}
\eeq

\subsection{Dirac delta behavior of the distribution $\delta_q(x)$}
\begin{figure}[htp]
\centering
\includegraphics[width=0.5\textwidth,keepaspectratio]{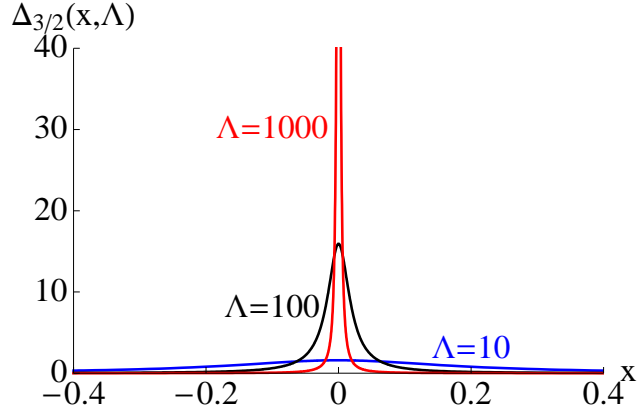}
\caption{Plot of $\Delta_{3/2}(x,\Lambda)$ for different values of $\Lambda$. Similar results are obtained for any value of $q\in]1,2[$.}
\label{fnascentqdelta}
\end{figure}

Let us define the following distribution 
\beq\label{nascentqdelta}
\Delta_q(x,\Lambda)\equiv\frac{\sin\left\{\frac{2-q}{q-1}\arctan[(q-1)\Lambda x]\right\}}{\pi x\left[1+(q-1)^2\Lambda^2x^2\right]^\frac{2-q}{2(q-1)}}\,\text{ with }q\in]1,2[ \,,
\eeq
which is related to $\delta_q(x)$ through
\beq
\delta_q(x)=\lim_{\Lambda\to\infty}\Delta_q(x,\Lambda)\,.
\eeq
The plot of such a distribution (see Fig. \ref{fnascentqdelta}) indicates that in the limit $\Lambda\to\infty$, $\Delta_q(x,\Lambda)$ will present a divergence at the origin and will be zero for all $x\not=0$, i.e., at first glance, $\delta_q(x)$ appears to be a representation of Dirac delta.

Let us now consider an analytic function, $f:\text{dom}\,f\subset\mathbb{R}\to\mathbb{C}$, which can be expanded in Taylor series around the origin such that the expression
\beq
f(x)=\sum_{k=0}^{\infty}\frac{f^{(k)}(0)}{k!}x^k
\eeq
is valid for all $x\in\text{dom}\,f$. Then we have
\beq \label{finite}
\int_{-\infty}^\infty f(x)\delta_q(x)\,dx=\int_{\text{dom}\,f} f(x)\delta_q(x)\,dx\,.
\eeq
Replacing $f(x)$ by its Taylor series, this expression yields 
\beq\label{intTaylorprev}
\int_{\text{dom}\,f} f(x)\delta_q(x)\,dx=\lim_{\Lambda\to\infty}\sum_{k=0}^{\infty}\frac{f^{(k)}(0)}{k!}\int_{\text{dom}\,f}\frac{x^{k-1}\sin\left\{\frac{2-q}{q-1}\arctan[(q-1)\Lambda x]\right\}}{\pi\left[1+(q-1)^2\Lambda^2x^2\right]^\frac{2-q}{2(q-1)}}\,dx\,,
\eeq
in which we must remark that $q$ belongs to the interval $]1,2[$. If $\text{dom}\,f$ is a bounded interval of $\mathbb{R}$, i.e., $\text{dom}\,f=]a,b[$, with $a<b$, then using the change of variables $z=(q-1)\Lambda x$, we obtain
\beq\label{intTaylor}
\int_{\text{dom}\,f} f(x)\delta_q(x)\,dx=\lim_{\Lambda\to\infty}\sum_{k=0}^{\infty}\frac{f^{(k)}(0)}{(q-1)^k\Lambda^kk!}\int_{(q-1)\Lambda a}^{(q-1)\Lambda b}\frac{z^{k-1}\sin\left(\frac{2-q}{q-1}\arctan z\right)}{\pi\left(1+z^2\right)^\frac{2-q}{2(q-1)}}\,dz\,.
\eeq

The first term of the sum that appears above is
\beq\label{firstterm}
f(0)\lim_{\Lambda\to\infty}\int_{(q-1)\Lambda a}^{(q-1)\Lambda b}\frac{\sin\left(\frac{2-q}{q-1}\arctan z\right)}{\pi z\left(1+z^2\right)^\frac{2-q}{2(q-1)}}\,dz\,.
\eeq
If $0<a<b$ or $a<b<0$, we straightforwardly see that expression above is equal to zero. If $a<0<b$, then using relation \eqref{piint} we obtain that expression \eqref{firstterm} is equal to $f(0)$. Finally, if we have either $a=0$ or $b=0$ (with $a<b$), then, also using relation \eqref{piint} we obtain that expression \eqref{firstterm} is equal to $f(0)/2$.

\begin{figure}[htp]
\centering
\includegraphics[width=0.5\textwidth,keepaspectratio]{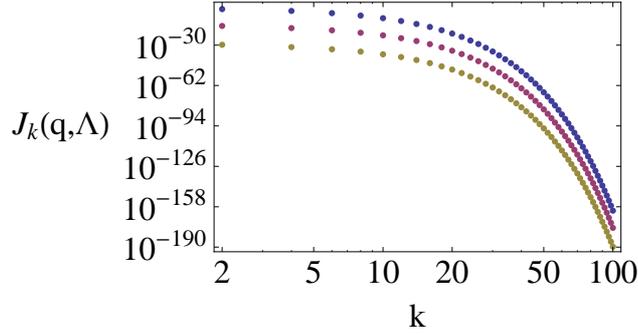}
\caption{The $k$-dependence of $J_k(q,\Lambda)$, numerically obtained, considering $b=-a=1$, for $q=1.4$, and different values of $\Lambda$. From top to bottom, $\Lambda=10$, $\Lambda=10^{10}$, $\Lambda=10^{20}$.}
\label{fratio1}
\end{figure}

In order to analyze the next terms of the sum given in \eqref{intTaylor}, let us first rewrite them as
\beq
\lim_{\Lambda\to\infty}f^{(k)}(0)J_k(\Lambda)\,\text{ with }k\in\mathbb{N}\text{ and }k>1\,,
\eeq
where
\beq
J_k(q,\Lambda)\equiv\frac{1}{(q-1)^k\Lambda^kk!}\int_{(q-1)\Lambda a}^{(q-1)\Lambda b}\frac{z^{k-1}\sin\left(\frac{2-q}{q-1}\arctan z\right)}{\pi\left(1+z^2\right)^\frac{2-q}{2(q-1)}}\,dz\,\text{ with }k\in\mathbb{N}\text{ and }k>1\,.
\eeq
$J_k(q,\Lambda)$ is a rapidly decreasing function of $k$ (see Fig. \ref{fratio1}), which makes the sum given in \eqref{intTaylor} converge, consistently with the finiteness of the domain of $f$ in integral in Eq. (\ref{finite}).  Moreover, from Fig. \ref{fratio1}, we can infer that, in the limit $\Lambda\to\infty$, $J_k(q,\Lambda)\to0$. Therefore, Eq. \eqref{intTaylor} implies
\beq\label{qdeltavalidity}
\int_a^b f(x)\delta_q(x)\,dx=\left\{\begin{array}{cc}
f(0)&\text{if }a<0<b\\
f(0)/2&\text{if either }a=0<b\text{ or }a<0=b\\
0&\text{if }0\not\in\,]a,b[.
\end{array}\right.\eeq

In the case when $\text{dom}\,f$ is unbounded, i.e. if $\text{dom}\,f=]a,\infty[$, or $\text{dom}\,f=]-\infty,b[$, or $\text{dom}\,f=\mathbb{R}$, a similar analysis yields once again relation \eqref{qdeltavalidity}. Moreover, we numerically tested the validity of the mentioned relation using some types of functions and distributions (for example the Gaussian and the Lorentzian). Hence it seems reasonable to conjecture that, for a wide class of functions, $\delta_q(x)$ indeed is a representation of Dirac delta. Thus, we can finally write
\beq \label{final}
\delta(x-x')=\frac{2-q}{2\pi}\int_{-\infty}^\infty e_q^{-i\xi(x-x')}\,d\xi \;\;\;(q\in[1,2[) \,.
\eeq

\section{Square integrability of $q$-plane waves}
\begin{figure}[htp]
\centering
\includegraphics[width=0.5\textwidth,keepaspectratio]{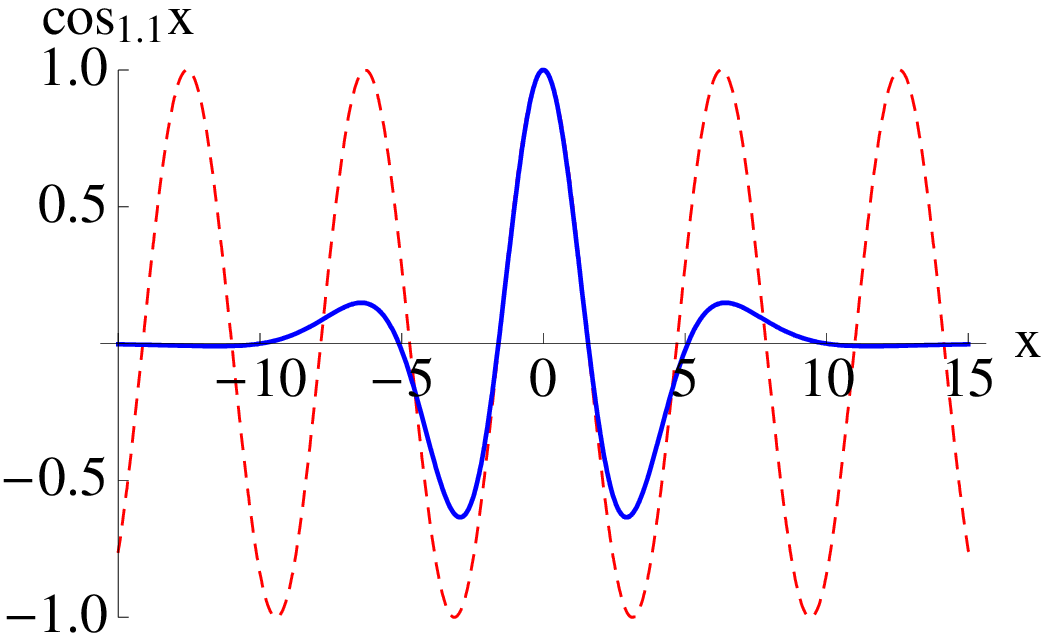}\vspace{1cm}
\includegraphics[width=0.5\textwidth,keepaspectratio]{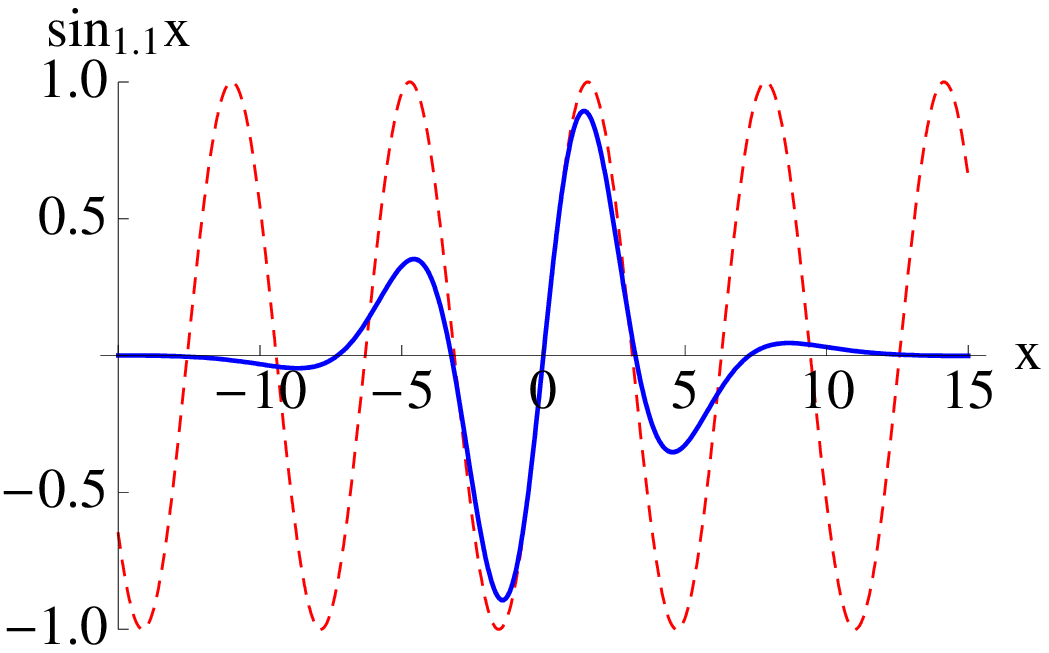}
\caption{{\it Top:} $\cos_{1.1}x$ (continuous curve) and $\cos x$ (dashed curve). {\it Bottom:} $\sin_{1.1}x$ (continuous) and $\sin x$ (dashed).
For $1 \le q<3$, $\cos_q x$ ($\sin_q x$) is an even (odd) function of $x$. For $1<q<3$, both functions $\cos_{q}x$ and $\sin_{q}x$ quickly decay when $|x|\to\infty$, in contrast with $\cos x$ and $\sin x$.}
\label{fqtrig}
\end{figure}

Let us consider the following function:
\beq\label{qpwave}
\Psi(x)=Ne_q^{i\xi x}=N\left[\cos_q(\xi x)+i\sin_q(\xi x)\right]\,\text{ with }q\in]1,3[\text{ and }\xi\not=0\,,
\eeq
which can be interpreted as a stationary $q$-plane wave, where the $q$-generalized trigonometric functions are defined, for any $x\in\mathbb{R}$, by (see also \cite{Borges1998}):
\beq
\cos_qx\equiv\text{Re}\left(e_q^{ix}\right)=\frac{\cos\left\{\frac{1}{q-1}\arctan[(q-1)x]\right\}}{\left[1+(q-1)^2x^2\right]^\frac{1}{2(q-1)}}\,,
\eeq
and
\beq
\sin_qx\equiv\text{Im}\left(e_q^{ix}\right)=\frac{\sin\left\{\frac{1}{q-1}\arctan[(q-1)x]\right\}}{\left[1+(q-1)^2x^2\right]^\frac{1}{2(q-1)}}\,.
\eeq
We illustrate these functions in Fig. \ref{fqtrig}.

We will determine now the value of the constant $N$ using the normalization condition given by
\beq\label{normqpwave}
\int_{-\infty}^\infty \Psi^*(x)\Psi(x)\,dx=1\,.
\eeq
Thus, we have
\beq
\frac{1}{N^2}=\int_{-\infty}^\infty e_q^{-i\xi x}e_q^{i\xi x}\,dx\,,
\eeq
which, using the definition of the $q$-exponential function given in (\ref{def}), can be written as
\beq
\frac{1}{N^2}=\int_{-\infty}^\infty\frac{1}{\left[1+(q-1)^2\xi^2x^2\right]^\frac{1}{q-1}}\,dx\,.
\eeq
Using the change of variables $\tan\theta=(q-1)|\xi| x$, this relation yields 
\beq
\frac{1}{N^2}=\frac{1}{(q-1)|\xi|}\int_{-\pi/2}^{\pi/2}(\cos\theta)^\frac{4-2q}{q-1}\,dx\,.
\eeq
Therefore, we obtain that the normalization constant is given by
\beq
N=\left[\frac{(q-1)|\xi|\Gamma\left(\frac{1}{q-1}\right)}{\sqrt{\pi}\Gamma\left(\frac{3-q}{2(q-1)}\right)}\right]^\frac{1}{2}\,.
\eeq
Let us emphasize that the function $\psi(x)=e_1^{i\xi x}$ (plane wave) cannot be normalized, whereas $q$-plane waves, with $q\in]1,3[$, have a finite norm.

\begin{figure}[htp]
\centering
\includegraphics[width=0.5\textwidth,keepaspectratio]{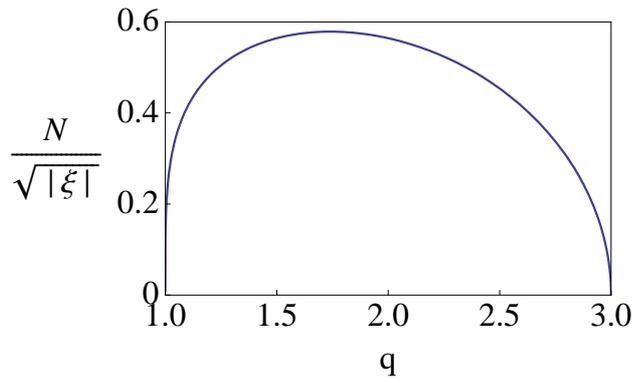}
\caption{The normalization constant $N$ as a function of $q$. $N$ goes to zero in the $q \to 1$ limit, thus recovering the well known non-normalizability of the plane waves.}
\end{figure}
\section{Conclusions}
From the analytical and numerical results shown in section II, we conjecture Eq. (\ref{final}), i.e., that $\delta_q(x)$ is indeed a generalization of the standard representation of Dirac delta in plane waves. Further research is welcome in order to establish which precise class of functions satisfy the relation \eqref{qdeltavalidity}.

Concomitantly, we found two new families of representations, namely expressions \eqref{pigamma} and \eqref{piint}, of the transcendental number $\pi$.  We tested the validity of such expressions for a set of values $n\in\mathbb{N}$ and $\alpha\in\mathbb{R}$. A demonstration is still required in order to formally establish these new families of representations of $\pi$.

A generalization of FT, namely, the so-called $q$-Fourier transform ($q$-FT) was developed in order to generalize the central limit theorem.  The possible analytic expression of the inverse $q$-FT remains to be found. It is known that, using the representation in plane waves of Dirac delta together with the expression of the direct FT, it is possible to find the expression of the inverse FT. Consequently, we suppose that the present $q$-generalization of the representation in plane waves of Dirac delta might be helpful in searching for an analytic expression of the inverse $q$-FT. Moreover, the present new representations of Dirac delta could be useful to handle some integrals that may appear in the analysis of certain physical phenomena.

Finally, we prove a physically appealing property, namely that the $q$-plane wave form $e_q^{ikx}$ is square-integrable (in other words, normalizable) for $1<q<3$, in contrast with the standard form, $e^{ikx}$, which is not. 

\acknowledgements
We acknowledge fruitful discussions with E.M.F. Curado, R.S. Mendes and F.D. Nobre, useful remarks by L.T. Cardoso, Y. Stein and C. Vignat, as well as partial support by Faperj and CNPq (Brazilian agencies).

\appendix*
\section{Trigonometric identity}
We establish here an expression for $\sin[2\alpha(q)\theta]$, with $2\alpha(q)\in\mathbb{N}$ and $\theta\in\mathbb{R}$, written in terms of $\sin\theta$ and $\cos\theta$. Firstly, we have
\beq
\sin[2\alpha(q)\theta]=\text{Im}\left[\left(\cos\theta+i\sin\theta\right)^{2\alpha(q)}\right],
\eeq
then, using binomial expansion we have
\beq
\sin[2\alpha(q)\theta]&=&\text{Im}\left[\sum_{k=0}^{2\alpha(q)}\binom{2\alpha(q)}{k}(\cos\theta)^{2\alpha(q)-k}(\sin\theta)^k(i)^k\right]\\
&=&-\sum_{\stackrel{k=1}{\text{(odd)}}}^{2\alpha(q)}(i)^{k+1}\binom{2\alpha(q)}{k}(\cos\theta)^{2\alpha(q)-k}(\sin\theta)^k.
\eeq
Finally,   this expression can be rewritten as follows:
\beq\label{trigidentity}
\sin[2\alpha(q)\theta]=\sum_{k=0}^{\left\lfloor\alpha(q)+\frac{1}{2}\right\rfloor-1}(-1)^k\binom{2\alpha(q)}{2k+1}(\cos\theta)^{2\alpha(q)-2k-1}(\sin\theta)^{2k+1}
\eeq
where we have used the {\it floor function} $\lfloor\,\,\rfloor$, defined, for any real number $x$, by $\lfloor  x \rfloor=n$ such that $n \le x<n+1$,  with $n \in \mathbb{Z}$.

\end{document}